# Variable Rate Video Compression using a Hybrid Recurrent Convolutional Learning Framework

Aishwarya Jadhav

Veermata Jijabai Technological Institute (VJTI)

anjadhav_b14@ce.vjti.ac.in

*Abstract*— In recent years, neural network-based image compression techniques have been able to outperform traditional codecs and have opened the gates for the development of learning-based video codecs. However, to take advantage of the high temporal correlation in videos, more sophisticated architectures need to be employed. This paper presents PredEncoder, a hybrid video compression framework based on the concept of predictive auto-encoding that models the temporal correlations between consecutive video frames using a prediction network which is then combined with a progressive encoder network to exploit the spatial redundancies. A variable-rate block encoding scheme has been proposed in the paper that leads to remarkably high quality to bit-rate ratios. By joint training and fine-tuning of this hybrid architecture, PredEncoder has been able to gain significant improvement over the MPEG-4 codec and has achieved bit-rate savings over the H.264 codec in the low to medium bit-rate range for HD videos and comparable results over most bit-rates for non-HD videos. This paper serves to demonstrate how neural architectures can be leveraged to perform at par with the highly optimized traditional methodologies in the video compression domain.

Keywords-video compression; neural networks; PredEncoder

## I. INTRODUCTION

A large part of the internet is glutted with video data, given the huge amount of videos uploaded onto and streamed from sites hosting video content. It has been estimated that videos account for about 80% of the overall traffic on the internet. YouTube alone makes up 37% of the total mobile internet traffic. All these numbers are given the fact that various compression techniques have been applied to video data before transmitting it across the web. Furthermore, along with the massive bandwidth requirements, videos are demanding in terms of storage requirements as well, taking up significantly more space than image data. Hence, it is indispensable for them to be compressed as much as possible, without drastically affecting their quality. As a result, any advancement in video compression technology can only help with the network bandwidth and storage burden of video data.

Since videos have spatial redundancy in each of the frames, as in the case of images, some of the techniques of image compression can be extended to video compression. The JPEG standard for image compression is based on block DCT transform or discrete wavelet transform (for the newer versions). The H.261 standard for video compression also makes use of DCT to reduce spatial redundancies. The recent trend in image compression techniques has been towards learning-based approaches. One work [1], employing Recurrent Neural Networks has even achieved superior performance as compared to the JPEG standard. These techniques can also be leveraged for video compression.

Apart from spatial redundancies, videos have a high degree of temporal redundancy between consecutive frames. This has already been exploited by almost all video codecs available today. For example, motion compensation provides a great way to capture temporal relationships between two consecutive frames. However, this technique involves the transmission of motion vectors along with any additional bits that capture the difference between the original frame and the motion-compensated frame. A reduction in the numbers of bits to be transmitted can be obtained if a compensated frame or base frame were available at the receiver end without the need of transmitting the motion vectors. In that case, the only bits that need to be passed will be for the difference between this base frame and the original frame.

This paper proposes 'PredEncoder' which is based on the aforementioned idea. PredEncoder employs a hybrid architecture that aims to model the spatio-temporal correlations in a video using a prediction framework to predict the next frames in a video sequence, supplemented by an iterative auto-encoder that encodes the differences between the predicted and the actual frames. By doing this, the framework effectively cuts down on the bits that are required to transmit data about the base frame by enabling the architecture to generate the base frames at the receiving end itself, in case of video transmission, or on the go, in case of storage and retrieval. Only the bits that represent the compact encodings of the difference images need to be transmitted or stored. This hybrid model need only be trained once and can be used with videos of any dimensions or frame rates. Furthermore, PredEncoder provides customizability in terms of quality vs bit-rate without the need to re-train the network. This is achieved by a progressive encoder that provides better quality at the cost of increased bit-rate.

## II. RELATED WORK

Much effort has been put into developing learning-based image compression schemes that either supplant some components of a traditional codec with neural networks, for optimization purposes, or aim to build a fully neural network based compression scheme. Jiang, in 1999, provided a comprehensive survey on this [7]. Since then, there have been numerous other efforts in this direction; more recently, Toderici et al. developed a novel architecture using recurrent neural networks in an auto encoder-decoder form for the compression of low resolution images [2]. In [1], they extended this scheme to the compression of images with a higher resolution using a block-based scheme. This architecture was able to achieve state-of-the-art results and provided a new direction for the development of learning-based encoders.

The domain of video compression has seen fewer efforts that involve a fully learning-based framework. One scheme [4] that was able to achieve significant results for a learning-based video codec uses the concept of motion extension for the prediction of blocks in a frame and then encodes the differences with Toderici's image compressor [1]. However, this approach is not tuned for providing variable compression rates for a block; it uses either 0 or 8 iterations for encoding each block. Another attempt [5] at a complete deep learning based video codec involves the use of optical flow estimation for prediction of next frames followed by a residual encoder [1]; this, however, involves very high computational costs. Our paper presents PredEncoder, which aims to describe an attempt at video compression using a fully learning-based architecture that employs variable rate compression for individual blocks in a video.

### III. THE PREDENCODER MODEL

PredEncoder is based on the concept of predictive coding which is used in many other image and video codecs. Predictive coding leverages the high degree of correlation, usually found between neighboring samples in a sequence, to predict or estimate the next-in-line sample, aiming to reduce the difference between the estimated and the actual sample. In the case of video compression, temporal correlations can be leveraged and the next-in-line video frame can be predicted based on the neighboring (previous) frames. The difference between the predicted frame and the actual frame can then be stored or transmitted. As depicted by the plots in Fig. 1, the difference frames have much less variance in terms of pixel values as compared to the original frames and hence, they can be represented using fewer bits. As the predictor accuracy increases, the variance in the difference frames decreases and consequently, the compression rates increase. Different image and video compression schemes employ varied predictors that aim to optimize the trade-off between accuracy and computational complexity.

#### A. Prediction Module

While statistical predictors have done well for the traditional codecs, they cannot beat the accuracy offered by learning-based predictors. The deep learning based video frame predictors available today try to predict future m time step frames based on the previous n time steps [8][9][10]. However, most of these predictors are computationally too complex to be employed for the compression of videos for transmission and storage as these will have to be run every time a video needs to be streamed over the internet or retrieved from storage. Moreover, these predictors will need to be run for every frame of the video at both, the encoder as well as the decoder end. Furthermore, we have a greater amount of information than that required by most of these predictors for their predictions, we have the actual frames that have to be predicted, at least at the encoder end. PredEncoder takes these advantages into consideration in order to reduce the computational complexity of its predictor.

Our predictor module is based on the next frame predictor by Lotter et al. [3] which works with frames of variable sizes. This model predicts the next 1 frame based on previous frames and hence, is computationally less taxing than other models for n-step prediction. The model stacks L modules consisting of Convolutional Recurrent Neural Network blocks, each of which tries to predict the input to it. The predictions are made based on states from the previous frames as well as from the layers above i.e. l+1, as depicted in Fig. 2. The errors between the predictions from and inputs to each of the modules are propagated as inputs to the l+1 level module. The level-0 module receives the original frame as input ($A$) and attempts to make a prediction of it ($\hat{A}$).

However, with this prediction model, at least one previous frame is required to make a prediction, the more the better. For example, the t+1 prediction will be better with t = 5 past actual frames as compared to t = 2 past frames. However, making too many actual frames available at the decoder end reduces the compression rate. To remediate this, we have modified the model to make predictions based on reconstructed frames as input rather than actual frames. This modification is based on the observation that the model also works well for predicting t+n frames using the past predicted frames as input in place of the actual frames, albeit with degraded prediction quality[3]. Using reconstructed frames provides better results than using predicted frames as input since the reconstructed frames contain additional information from the Encoder.

#### B. Encoder

The encoder is used mainly for compressing the differences between the predicted frames and the actual frames. It is also used for encoding the first frames of the video sequences since these first frames cannot be predicted. The compressed encodings are then sent across the network or stored. For these purposes, we utilize the progressive encoding scheme proposed by Toderici et al in [1]. This block-based encoder framework consists of an n-layer ConvLSTM architecture that acts as the encoder, a binarizer that is the bottleneck for the framework and a ConvLSTM based decoder. It tries to produce a compact representation of an image, or, of a difference image. For a 32x32 image block, one iteration of a single frame produces a code of size (2x2x32) which yields a compression ratio of 1:192. The encoder progressively encodes the differences, with each of the iterations adding more bits to the representation, which in turn improves the quality of the decoded image.

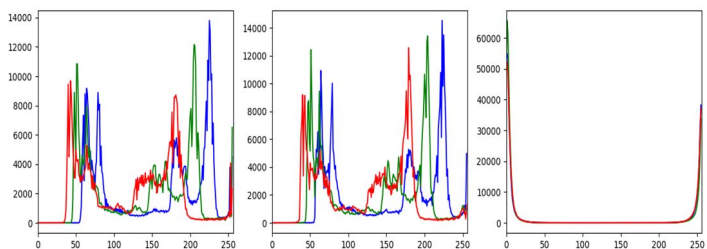

**Fig. 1:** (a). Histogram of original frame pixel values. (b). Histogram of predicted frame pixel values. (c) Histogram of difference frame pixel values

This progressive nature of the encoder has been leveraged in our architecture to achieve variable rate compression. We iterate over the image, that is input to the encoder, only until the difference between the decoded and the input image is greater than a threshold value of the quality metric MAE (mean absolute error). The maximum number of iterations required for any difference image has been empirically set to 8. Therefore, only 3 additional bits will need to be sent with each block, denoting the number of iterations with which it was encoded. Moreover, since the first image of each sequence has the maximum variance, as it is not a difference image, it is always encoded with 16 iterations, as determined empirically.

### C. Entropy Coding

We use the progressive learning-based entropy coding scheme [1] on the binarized codes from the Encoder. Since the focus of this paper is a learning-based video encoder-decoder system, we do not provide any specific optimizations to the entropy coder and just employ it as an additional compression mechanism after the encodings are generated.

## IV. TRAINING

### A. Dataset

We picked a selection of 2 million Youtube videos from 45 different categories. Youtube contains a wide mix of videos shot in various settings such as with a car-mounted camera, or in a professional setting or videos that are parts of video game clips. All these videos contain varying degrees of object and camera movements. This provides a rather tough dataset for the model to learn on. We downloaded the 30 fps color videos of the best resolution (1280x720). The video frames were resized to 960x512. Both the predictor and the encoder have been trained on this dataset.

### B. Training for Predictor

1.2 million 10-frame sequences were randomly taken from the video dataset for training. A separate set of 400 videos was used as the validation set. The predictor was trained for 15 epochs. For this initial training, original frames were used as input for each prediction. This was to enable the model to learn the basic characteristics of videos.

### C. Training for Encoder

We randomly cropped 8 million 32x32 RGB image blocks from random frames in the videos. The encoder model was first trained for 10 epochs on this image dataset. Then, we created another dataset consisting of differences images by

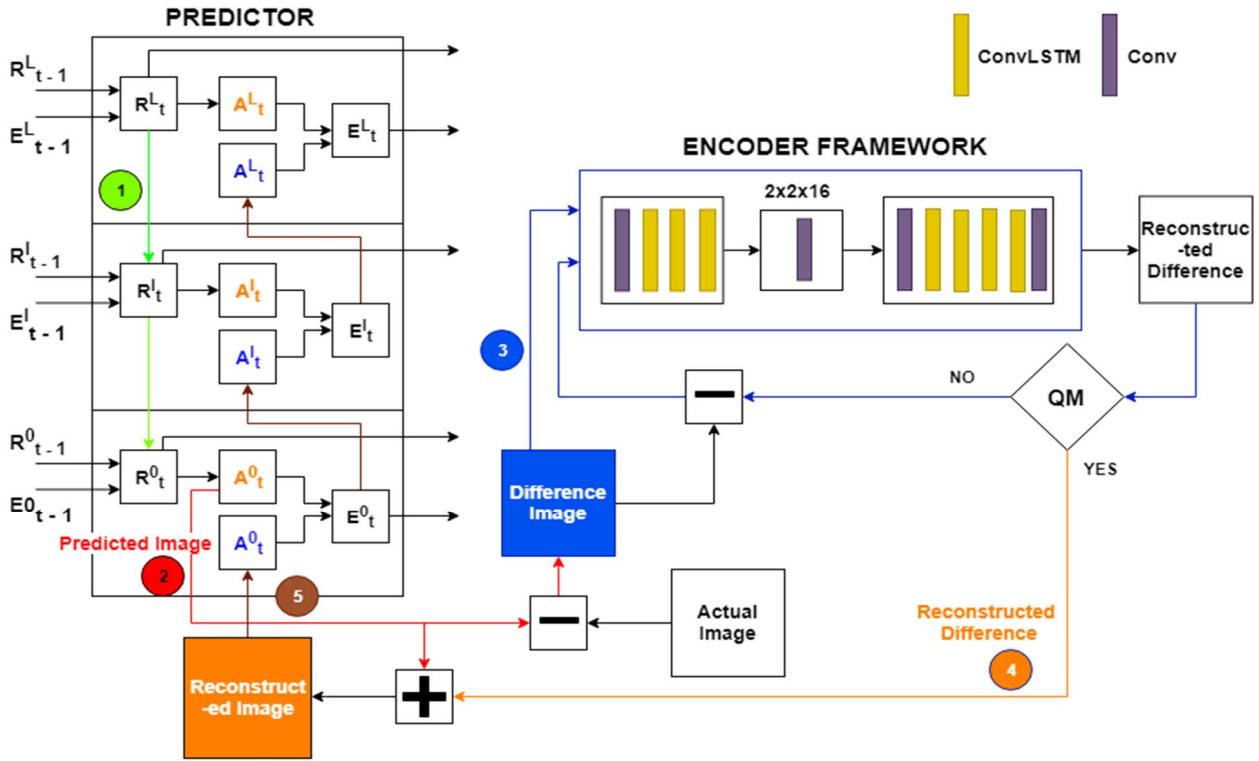

**Fig. 2:** A block diagram of PredEncoder. The colored arrows denote the varies steps in the encoding of the $t^{th}$ frame. The numbers next to the arrows denote the sequence. First comes the downward pass which updates the states of R. These R nodes are then used in the prediction of $A_t^l$ in step 5. However, the prediction from $R_t^0$ i.e. $A_t^0$ acts as the predicted image in step 2. The difference between this predicted frame and the actual frame is fed to the encoder frame, which iteratively produces the reconstructed difference in step 3. In step 4, the quality of this reconstructed difference is checked using the Quality Metric (QM), and if greater, it is added to the predicted image from step 2 to form the reconstructed frame. This is then fed to the predictor in place of the actual image, which is unavailable at the decoder end, for the upward pass in step 5. The equations describing the modules in this architecture are given in fig. 3

$$A_t^l = maxpool(relu(conv(E_t^{l-1})))$$

$$\widehat{A_t^l} = Relu\left(Conv(R_t^l)\right)$$

$$E_t^l = \left[Relu\left(A_t^l - \widehat{A_t^l}\right); Relu\left(\widehat{A_t^l} - A_t^l\right)\right]$$

$$R_t^l = ConvLSTM(E_{t-1}^l, R_{t-1}^l, R_t^{l+1})$$

**Figure 3:** Equations of modules used in fig. 2

running the predictor model on the video dataset and taking the difference of the predicted and actual frames. We sampled 2 million difference images of size 32x32 according to a normal distribution of their pixel value averages. This was done to avoid including a lot of 0-difference images in the new dataset. We fine-tuned the encoder model to compress the difference images well. Further, we trained the entropy coder jointly with the trained encoder on a subset of the dataset used to train the encoder.

### D. Joint Training

Finally, we plugged the two models together to jointly train them. This joint training fine-tuned the predictor to make predictions using reconstructed frames as opposed to using actual frames. Since a better reconstruction of the difference images by the encoder framework affects the prediction quality of the predictor and a better prediction by the predictor allows for a lower variance in the difference images and hence, a better reconstruction by the encoder, both the models are fine-tuned to the compression by this joint training. We do this training using the video dataset used for pre-training the predictor along with the same validation set. For this, 10 epochs were used.

For each of the trainings, we used the Adam optimizer and began with a learning rate of 0.001, reducing it by a factor of 2 after every 4 epochs. Also, we used 16 iterations for the first frame and 8 (fixed) iterations for the difference frames during the training phase. Variable iterations and the 3-bit counts are not employed during training. Entropy coder is not used in the joint training.

### E. Optimization Function

For the predictor training, MAE between the predicted and the actual frames averaged over all frames in a video sequence is used as the optimization parameter:

$$L_{pred} = \frac{1}{N} \times \sum_N \frac{1}{F-1} \times \sum_{i=0}^{F-2} |a_i - p_i|$$

where N is the batch size and F is the number of frames in a video. $a_i$ is the $i^{th}$ actual frame and $p_i$ is the $i^{th}$ predicted frame. We do not consider the first frame in the calculation of the MAE loss for the predictor.

For the encoder training, MAE between the reconstructed differences and actual differences averaged over all the iterations for all the blocks of a particular difference image has been optimized:

$$L_{decoded} = \frac{1}{N} \times \frac{1}{F} \times \sum_N \sum_F \frac{1}{B} \sum_{j=0}^{B-1} \frac{1}{T} \times \sum_{i=0}^{T-1} |b_i^j - d_i^j|$$

where N is the batch size, F is the number of frames in the video, T is the number of iterations required for the reconstruction of a block. $b_i^j$ denotes the $i^{th}$ iteration input of the $j^{th}$ block, to the encoder framework; whereas $d_i^j$ denotes the output for the $j^{th}$ block in the $i^{th}$ iteration.

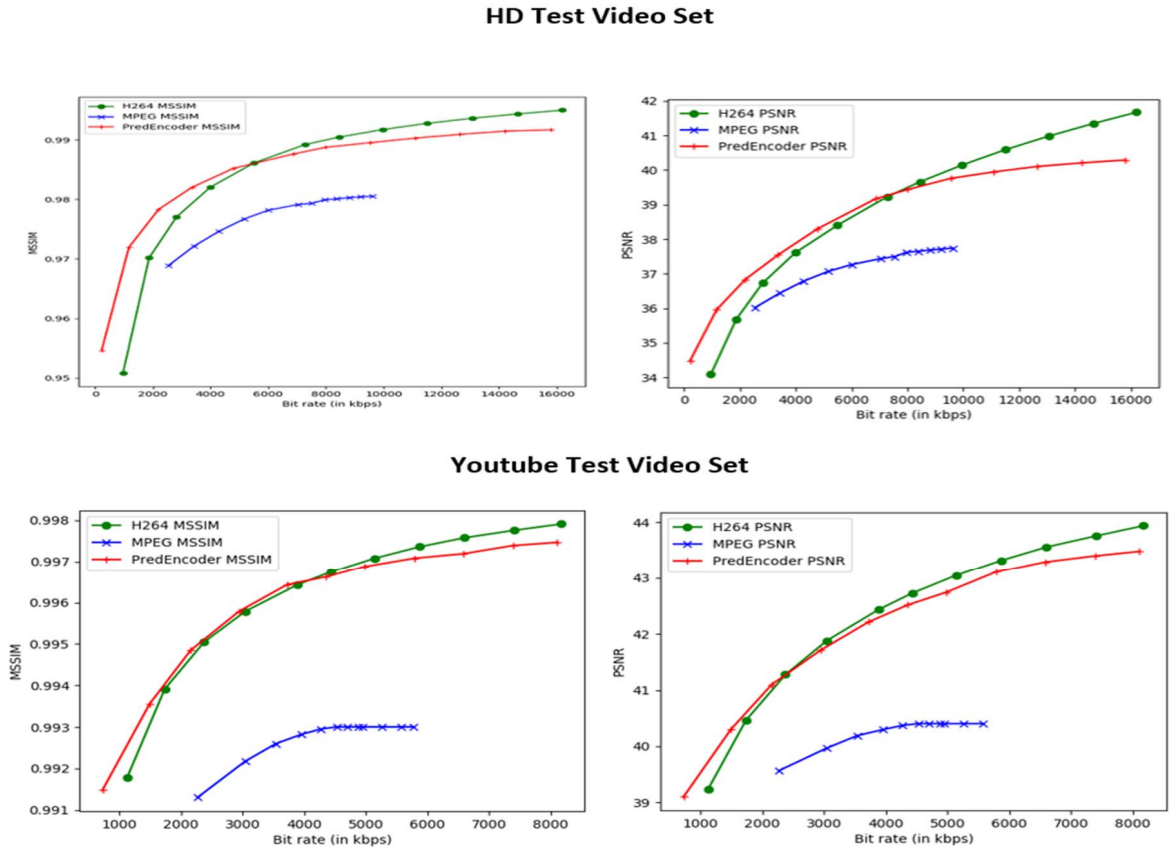

**Figure 4:** Left: MSSIM vs bit-rate plots; Right: PSNR-HVS vs bit-rate plots

For the joint training, the sum of the losses of the predictor and the encoder was used as the optimization metric:

$$L_{reconstructed} = L_{predicted} + L_{decoded}$$

## V. EXPERIMENTAL RESULTS

We performed tests on two kinds of video sets. The first set consisted of 8 HD videos from [6]. Each of these videos was in the YUV 4:2:0 8-bit depth raw video format. The videos of size 1920 x1080 were downsized to 960x512 and converted to a 30fps frame rate. This set represents professionally shot videos that inherently have a very high bit-rate, each raw video being about 2 GB in size. There is usually a high amount of redundancy in such videos. The other set on which we tested our system consisted of 8 videos downloaded from Youtube, each with size 1280x720, which were also downsized to 960x512. These videos represent the bulk of videos on the internet today. Along with the movement of objects, there is a lot of random camera movement involved in these videos. So, the temporal relations between consecutive frames are much more difficult to capture. All the 16 test videos from both categories are stored in a raw .avi format.

We first reconstruct all the test videos using PredEncoder with different bit-rates controlled by the quality metric MAE, which determines the number of iterations required for a particular block. During this encoding, we observe that the about 8% of the blocks, on an average, required 0 iterations with the most stringent MAE requirement or with the highest bit-rate, while the average was 42% for the lowest bit-rate that we experimented with. The 3-bit metadata that indicates the number of iterations for a given block just accounts for ~2.5% of the bit-stream but it leads to a bit-rate saving of 20% - 65% over the method which uses a fixed number of iterations (8 in our experiments).

For the baseline with which to compare our model, we select two traditional video codecs: MPEG-4 and H.264, which are amongst the most popular ones used today. We encode the raw test sequences from both the test sets with various bit-rates using these codecs and compare the reconstructed video frames to the reconstructed frames from PredEncoder with equivalent bit-rates. The frames from these three codecs are evaluated using the PSNR-HVS[12] and MS-SSIM [11] metrics. These metrics are calculated between the individual corresponding frames from the original and reconstructed videos, averaged over all the frames in the video sequence. The results are then

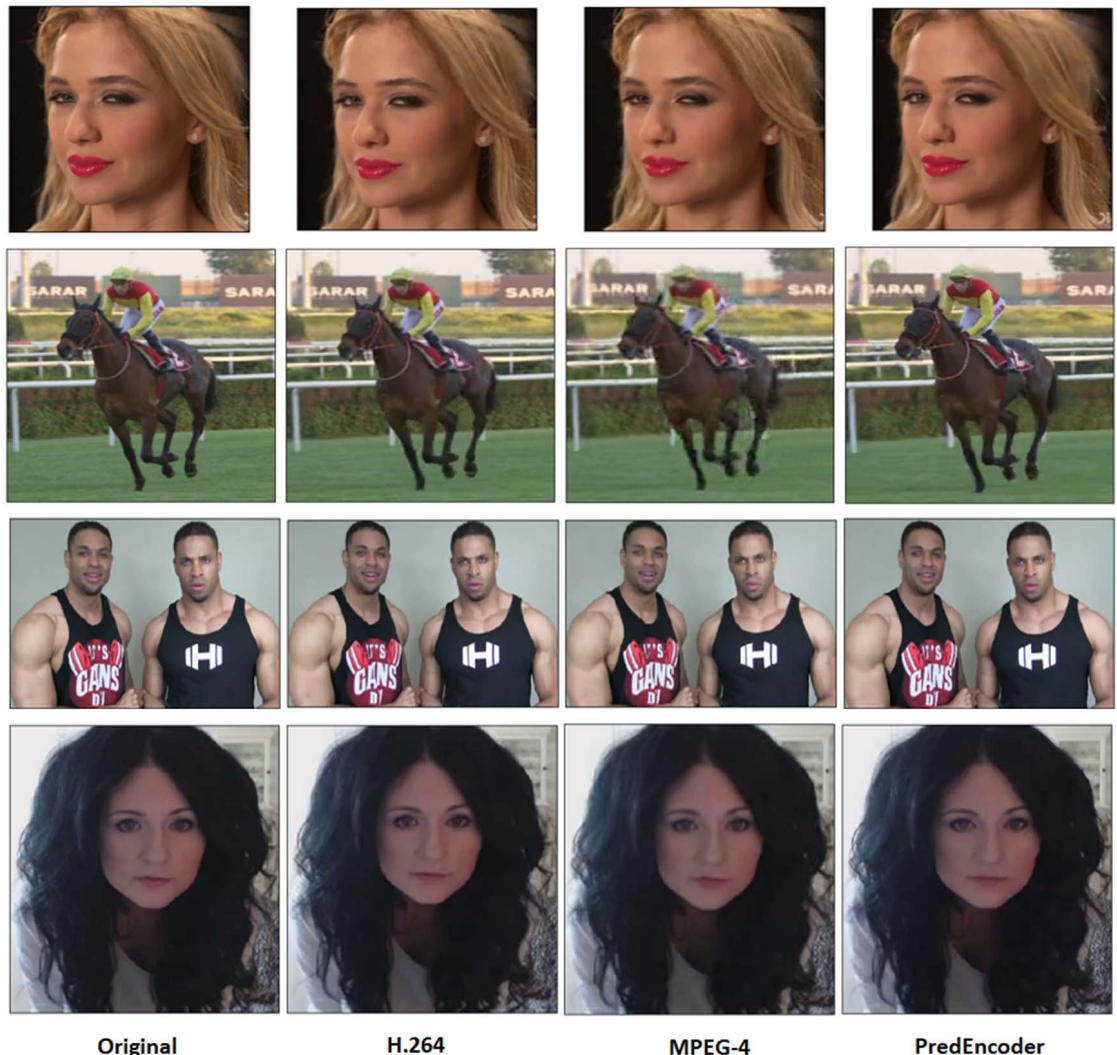

**Figure 3:** Sample frames from videos encoded and decoded using different video codecs at equivalent bit-rates

averaged separately over each of the two test sets and are as reported in Fig 4.

As can be seen from the plots in Fig. 4, since there is a high amount of redundancy in HD videos, PredEncoder is able to achieve quality comparable to that of H.264 codec with a lower bit-rate. Almost 38% of the blocks in the videos in this bit-rate range required fewer than 2 iterations each. We observe that after a certain threshold, the quality of PredEncoder reconstructed videos increases marginally with the increase of bit-rate. However, after this quality threshold, the differences between the original and the reconstructed frames are hard to spot by the perceptual vision, much less at a 30fps frequency. This can be observed in the samples of Fig. 5 which depict original and reconstructed frames with comparable bit-rates in the range >7000 Kbps where H.264 normally outperforms PredEncoder. 4.

We further observe, in the plots of Youtube videos in Fig. 4, that due to the random camera movements, the bit rate savings over H.264 codec are lesser than those for HD videos. Moreover, since the quality of Youtube videos is generally lower than that of HD videos, PredEncoder is able to perform almost comparable to H.264 even in the higher bitrate range. The perceptive differences are again barely visible in Fig. 5 sample frames.

## VI. DISCUSSION

Though not able to completely outperform the H.264 codec, PredEncoder has been able to achieve significant improvements over the MPEG4 codec and has been able to perform fairly well in comparison to H.264 under some circumstances. The only drawback is that quality vs. bit-rate ratio decreases with higher bit-rates so much so that it falls short of the quality levels achieved by H.264. However, as one of the earliest attempts at a fully learning-based video codec, PredEncoder offers the following salient advantages:

1. The frames of a video can be encoded and decoded sequentially as all the information needed to decode a frame is available from the previously decoded frames. It is also a block-based encoding scheme. So, videos of arbitrary length and size can be encoded using PredEncoder.
2. Significant bit rate reductions can be achieved with videos that are not HD, like most of the video traffic on the internet. Even with HD images, a bit-rate reduction is observed until a certain quality threshold. However, even beyond this threshold, the visible video quality is much better as compared to MPEG4 where block artifacts are visible in the result.
3. Around 10.89% of the reduction in bit rate, on an average, is due to the entropy encoder. There remains the scope of further exploring an entropy coder that can be optimized to work well with this codec to achieve an even higher compression rate.
4. The system provides the facility of customizing the quality metric that determines the number of iterations required for each block. Different domain-based fidelity metrics can be used in the design of the encoder and decoder to control the bit-rate.
5. There is just the 3-bit counter, that indicates the number of iterations for a particular block, that needs to be passed as metadata in the bit-stream of a video. This can also be reduced using a larger block size. For many traditional HEVC codecs, that provide several bit-rate and quality control options, more sophisticated metadata needs to be passed.

Being a learning-based methodology that employs deep neural network architectures, the computational complexity of this method is higher than most traditional codecs. This can be reduced using optimizations tuned to the system and hardware suited for use with deep learning models.

## VII. CONCLUSION

PredEncoder is one of the first end-to-end neural network-based video compression frameworks with components such as predictor, encoder, binarizer and entropy coder, all employing deep neural architectures to effectively model spatio-temporal correlations and develop a compact variable rate encoding for a video sequence. The paper, thus, presents experimental evidence of the power of modern deep learning frameworks.